\documentclass[a4paper,11pt]{article}
\usepackage{jinstpub} % for details on the use of the package, please see the JINST-author-manual
\usepackage{lineno}
\title{PICOSEC Micromegas Precise-timing Detectors: Development towards Large-Area and Integration}

\author[a,b,1]{Y. Meng\note{Corresponding author.}}
\author[c]{R. Aleksan}
\author[d]{Y. Angelis}
\author[e]{J. Bortfeldt}
\author[f]{F. Brunbauer}
\author[g,h]{M. Brunoldi}
\author[d]{E. Chatzianagnostou}
\author[i]{J. Datta}
\author[j]{K. Degmelt}
\author[k]{G. Fanourakis}
\author[g,h,1]{D. Fiorina\note{Now at Gran Sasso Science Institute, Viale F. Crispi, 7 67100 L’Aquila, Italy.}}
\author[f,l]{K. J. Floethner}
\author[l]{M. Gallinaro}
\author[m]{F. Garcia}
\author[c]{I. Giomataris}
\author[j]{K. Gnanvo}
\author[c,2]{F.J. Iguaz\note{Now at SOLEIL Synchrotron, L’Orme des Merisiers, Départementale 128, 91190 SaintAubin, France.}}
\author[f]{D. Janssens}
\author[c]{A. Kallitsopoulou}
\author[o]{M. Kovacic}
\author[j]{B. Kross}
\author[c]{P. Legou}
\author[a,b]{Z. Li}
\author[f,p]{M. Lisowska}
\author[a,b]{J. Liu}
\author[a,b]{Y. Ma}
\author[f,c,4]{I. Maniatis\note{Now at Department of Particle Physics and Astronomy, Weizmann Institute of Science, Rehovot, 7610001, Israel.}}
\author[j]{J. McKisson}
\author[f,q]{H. Muller}
\author[f]{E. Oliveri}
\author[f,r]{G. Orlandini}
\author[j]{A. Pandey}
\author[c]{T. Papaevangelou}
\author[s]{M. Pomorski}
\author[f]{L. Ropolewski}
\author[d,t]{D. Sampsonidis}
\author[f]{L. Scharenberg}
\author[f]{T. Schneider}
\author[c,5]{L. Sohl\note{Now at TÜV NORD EnSys GmbH Co. KG.}}
\author[f]{M. van Stenis}
\author[u]{Y. Tsipolitis}
\author[d,t]{S.E. Tzamarias}
\author[v]{A. Utrobicic}
\author[g,h]{I. Vai}
\author[f]{R. Veenhof}
\author[g,h]{P. Vitulo}
\author[a,b]{X. Wang}
\author[f,w]{S. White}
\author[j]{W. Xi}
\author[a,b]{Z. Zhang}
\author[a,b]{L. Zhao}
\author[a,b]{Y.Zhou}
\affiliation[a]{State Key Laboratory of Particle Detection and Electronics, University of Science and Technology of China, 230026, Hefei, China}
\affiliation[b]{Department of Modern Physics, University of Science and Technology of China, Hefei 230026, China}
\affiliation[c]{IRFU, CEA, Université Paris-Saclay, F-91191 Gif-sur-Yvette, France}
\affiliation[d]{Department of Physics, Aristotle University of Thessaloniki, University Campus, GR-54124, Thessaloniki, Greece}
\affiliation[e]{Department for Medical Physics, Ludwig Maximilian University of Munich, Am Coulombwall 1, 85748, Garching, Germany}
\affiliation[f]{European Organisation for Nuclear Research (CERN), CH-1211, Geneve 23, Switzerland}
\affiliation[g]{Dipartimento di Fisica, Università di Pavia, Via Bassi 6, 27100 Pavia (IT)}
\affiliation[h]{INFN Sezione di Pavia, Via Bassi 6, 27100 Pavia (IT)}
\affiliation[i]{Stony Brook University, Dept. of Physics and Astronomy, Stony Brook, NY 11794-3800, USA}
\affiliation[j]{Jefferson Lab, 12000 Jefferson Avenue, Newport News, VA 23606, USA}
\affiliation[k]{Institute of Nuclear and Particle Physics, NCSR Demokritos, GR-15341 Agia Paraskevi, Attiki, Greece}
\affiliation[l]{Helmholtz-Institut für Strahlen- und Kernphysik, University of Bonn, Nußallee 14–16, 53115 Bonn, Germany}
\affiliation[m]{Laboratório de Instrumentacão e Física Experimental de Partículas, Lisbon, Portugal}
\affiliation[n]{Helsinki Institute of Physics, University of Helsinki, FI-00014 Helsinki, Finland}
\affiliation[o]{Faculty of Electrical Engineering and Computing, University of Zagreb, 10000 Zagreb, Croatia}
\affiliation[p]{Université Paris-Saclay, F-91191 Gif-sur-Yvette, France}
\affiliation[q]{Physikalisches Institut, University of Bonn, Nußallee 12, 53115 Bonn, Germany}
\affiliation[r]{Friedrich-Alexander-Universität Erlangen-Nürnberg, Schloßplatz 4, 91054 Erlangen, Germany}
\affiliation[s]{CEA-LIST, Diamond Sensors Laboratory, CEA Saclay, F-91191 Gif-sur-Yvette, France}
\affiliation[t]{Center for Interdisciplinary Research and Innovation (CIRI-AUTH), Thessaloniki 57001, Greece}
\affiliation[u]{National Technical University of Athens, Athens, Greece}
\affiliation[v]{Ruđer Bošković Institute, Bĳenička cesta 54, 10000 Zagreb, Croatia}
\affiliation[w]{University of Virginia, VA, USA}

\emailAdd{mallory4869@mail.ustc.edu.cn}

\abstract{PICOSEC Micromegas (MM) is a precise timing gaseous detector based on a Cherenkov radiator coupled with a semi-transparent photocathode and an MM amplifying structure. The detector conceprt was successfully demonstrated through a single-channel prototype, achieving sub-25 ps time resolution with Minimum Ionizing Particles (MIPs).  A series of studies followed, aimed at developing robust, large-area, and scalable detectors with high time resolution, complemented by specialized fast-response readout electronics. This work presents recent advancements towards large-area resistive PICOSEC MM, including 10 $\times$ 10 $\text{cm}^2$ area prototypes and a 20 $\times$ 20 $\text{cm}^2$ prototype, which features the jointing of four photocathodes. The time resolution of these detector prototypes was tested during the test beam, achieved a timing performance of around 25 ps for individual pads in MIPs. Meanwhile, customized electronics have been developed dedicated to the high-precision time measurement of the large-area PICOSEC MM. The performance of the entire system was evaluated during the test beam, demonstrating its capability for large-area integration. These advancements highlight the potential of PICOSEC MM to meet the stringent requirements of future particle physics experiments.}


\begin{document}
\maketitle
\flushbottom

\section{Introduction}
\label{sec:intro}

Precise-timing detection techniques are in high demand for future particle physics experiments. PICOSEC Micromegas (MM) is a precise timing gaseous detector based on a Cherenkov radiator coupled to a semi-transparent photocathode and a Micromegas amplifying structure \cite{a}. A single-channel prototype has been successfully manufactured for proof of concept and achieved sub-25 ps time resolution with Minimum Ionizing Particles (MIPs), followed by ongoing developments toward detector optimization and large-area coverage. Recent advancements include a new single-channel structural design which achieved an improving timing performance of 12.5 ps \cite{UTROBICIC2025170127}, the development of robust photocathodes and resistive Micromegas, the exploration of alternative gas mixtures \cite{aime2024simulation}, and the extension of \textmu RWELL technology to the PICOSEC concept \cite{weisenberger2023murwell}. Expanding the detection area is crucial for the application in future particle physics experiments. While several approaches to multi-channel prototypes have been extensively explored and developed \cite{aune2021timing, lisowska2023sub}, large-area prototypes still face challenges related to robustness, uniformity, and the limitations imposed by the maximum size of the photocathode. In this paper, we present approaches to large-area expansion while ensuring robustness. The large-area resistive prototypes were designed and fabricated, and a series of tests were conducted with the prototypes to validate their performance. In addition, customized electronics systems dedicated to facilitating the fast signal processing of large-area PICOSEC MM were developed. The performance of the entire system has been tested through the beam test campaign.

%Recent advancements include the development of a new single-channel structure aimed at improving stability, ensuring signal integrity, and achieving uniform time response over the entire active area. The new design achieved an excellent time resolution of 12.5 ps within the pad's central region. Replacing the original CsI photocathode, which is vulnerable to ion backflow and sensitive to humidity, with more robust photocathodes is crucial. Promising alternatives include B4C, DLC, and nanodiamonds, with DLC photocathodes being deposited using the magnetron sputtering technique and having undergone extensive testing. Detectors with precise timing abilities and large detection areas, while operating under great counting rates and harsh radiation environments are facing crucial challenges that require further technical exploration. 

\section{Development of Multi-channel PICOSEC Micromegas Prototypes}

Based on the PICOSEC concept, multi-channel PICOSEC MM with an effective area of 10 $\times$ 10 $\text{cm}^2$ PICOSEC MM was designed, produced, and tested, achieving a time resolution of 25 ps in beam tests \cite{lisowska2023sub}. Extensive research has been conducted on the resistivity of Micromegas, and a 10 $\times$  10 $\text{cm}^2$ area resistive PICOSEC MM was developed by adding a resistive diamond-like carbon (DLC) layer on the Micromegas. Additionally, a double DLC layer was developed to further enhance charge evacuation and evaluate rate capability. Other approaches, such as the PICOSEC detector concept based on the \textmu RWELL structure, have also been explored for the 10 $\times$ 10 $\text{cm}^2$ area.

\begin{figure}[htbp]
\centering
    % 第一张图片
    \begin{minipage}[t]{0.45\textwidth} % 设置图片宽度为页面宽度的45%
        \centering % 图片和文字居中
        \includegraphics[width=\textwidth]{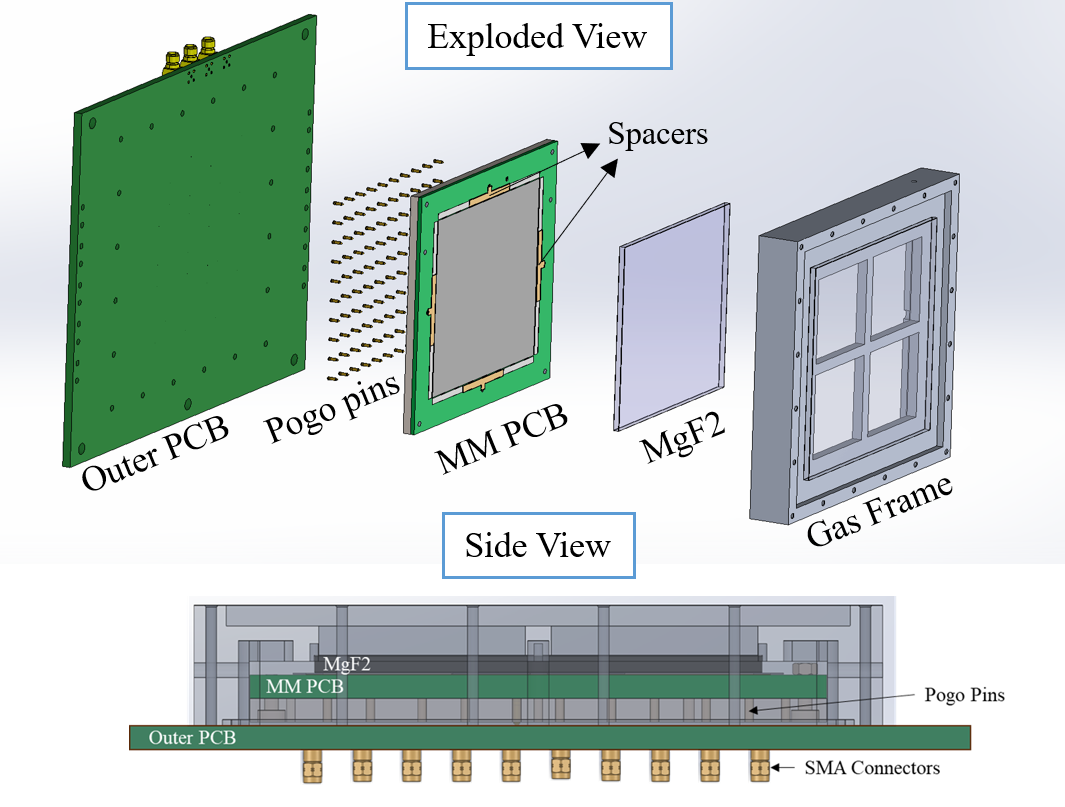} % 插入图片
        \vspace{2mm} % 图片与文字间的垂直间距
        \textbf{(a)} % 添加文字标注 (a)
    \end{minipage}
    \qquad
    % 第二张图片
    \begin{minipage}[t]{0.42\textwidth}
        \centering % 图片和文字居中
        \includegraphics[width=\textwidth]{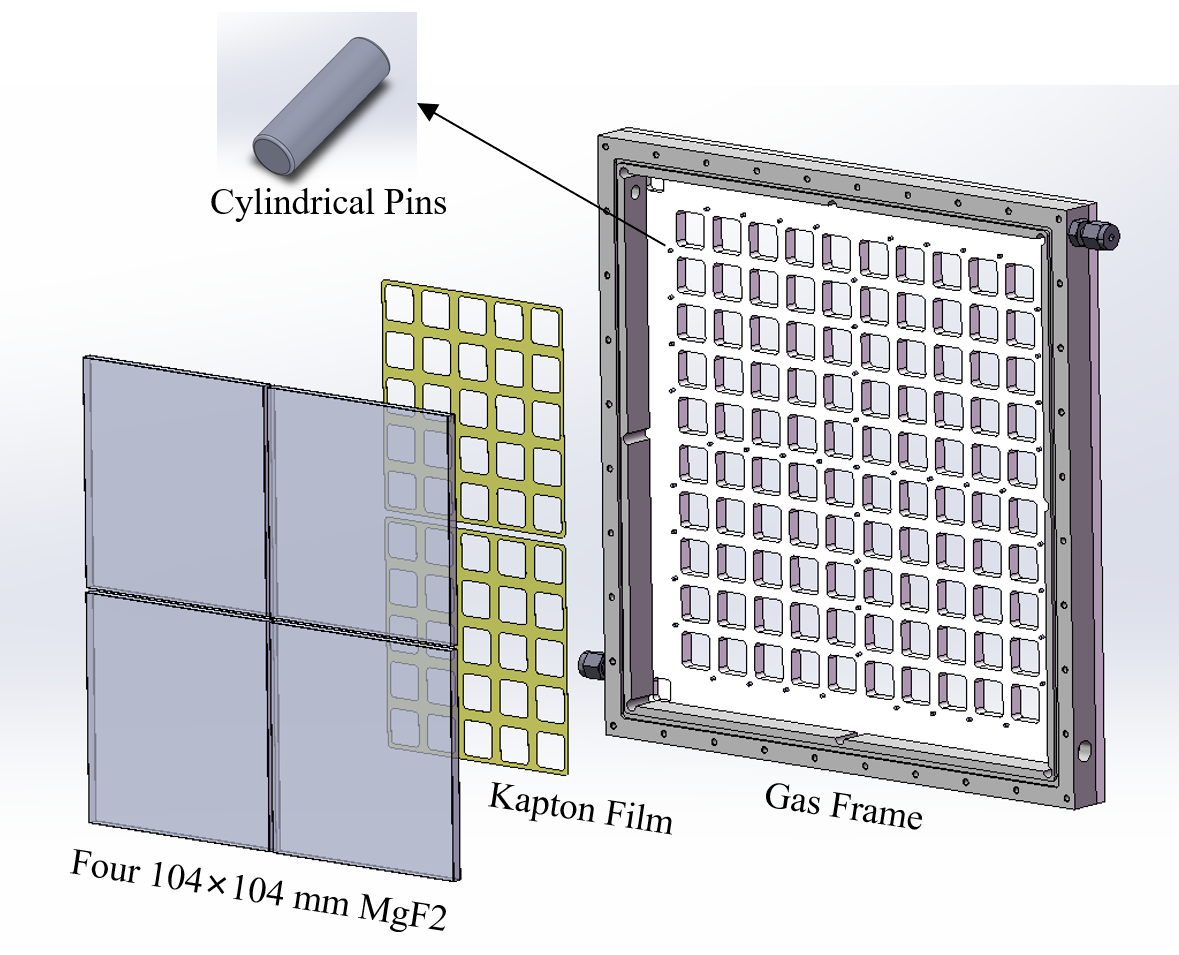} % 插入图片
        \vspace{2mm} % 图片与文字间的垂直间距
        \textbf{(b)} % 添加文字标注 (b)
    \end{minipage}
\caption{(a) Scheme of the 10 $\times$ 10 $\text{cm}^2$ PICOSEC MM layout. (b) Schematic diagram of the four crystals assembly in the 20 $\times$ 20 $\text{cm}^2$ PICOSEC MM. \label{fig:f1}}
\end{figure}

To enable the expansion to larger areas while ensuring robustness, a scalable detector scheme was proposed. A 10 $\times$ 10 $\text{cm}^2$ PICOSEC MM prototype incorporating 100 channels of 9.7 $\times$ 9.7 $\text{mm}^2$ pads was designed and manufactured. It consists of a gas frame with a quartz window, a whole 104 \text{mm} $\times$ 104 \text{mm} $\times$ 3 \text{mm} $\text{MgF}_2$ as Cherenkov radiator, an MM printed circuit board (PCB), and an outer PCB to extract signals, as shown in figure \ref{fig:f1} (a). A thermal bonding method was applied to fabricate the resistive Micromegas \cite{m} with a resistivity of approximately 50 M\(\Omega\)/sq. The amplification gap of the Micromegas is approximately 100 \textmu m, while the pre-amplification (PA) gap defined by the spacers are around 170 \textmu m. The MM board was adhesively bonded to a ceramic board to ensure its flatness and maintain detector uniformity. Based on a similar design, a 20 $\times$ 20 $\text{cm}^2$ prototype design was then developed, allowing the assembly of four crystals to extend the detector to an even larger area. Figure \ref{fig:f1} (b) illustrates the design concept for assembling the four 10 $\times$ 10 $\text{cm}^2$ $\text{MgF}_2$ crystals in the detector frame. The four crystals were placed directly on the frame using cylindrical pins for alignment. The spacers used here are approximately 190 \textmu m, serving to define the PA gap. Kapton films were inserted underneath the crystals to compensate for thickness variations. The Micromegas was designed as a single unit, which utilized the thermal bonding method for fabrication and was reinforced with a ceramic plate. This design and manufacturing approach ensures the scalability of the detector for larger areas while maintaining stability.

\section{Performance of the Large-area Prototypes}

\begin{figure}[htbp]
\centering
    % 第一张图片
    \begin{minipage}[t]{0.35\textwidth} % 设置图片宽度为页面宽度的45%
        \centering % 图片和文字居中
        \includegraphics[width=\textwidth]{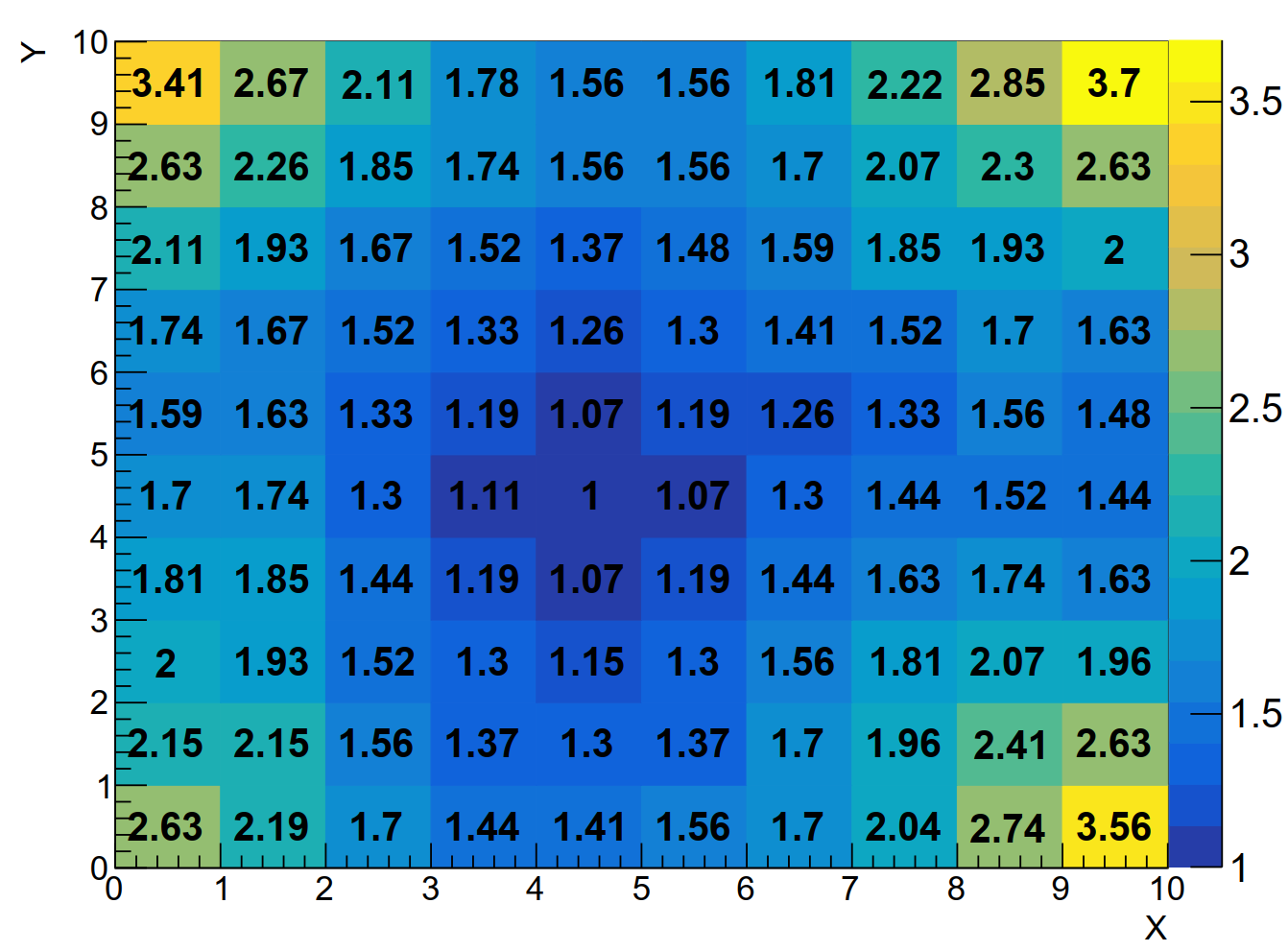} % 插入图片
        \vspace{2mm} % 图片与文字间的垂直间距
        \textbf{(a)} % 添加文字标注 (a)
    \end{minipage}
    \qquad
    % 第二张图片
    \begin{minipage}[t]{0.45\textwidth}
        \centering % 图片和文字居中
        \includegraphics[width=\textwidth]{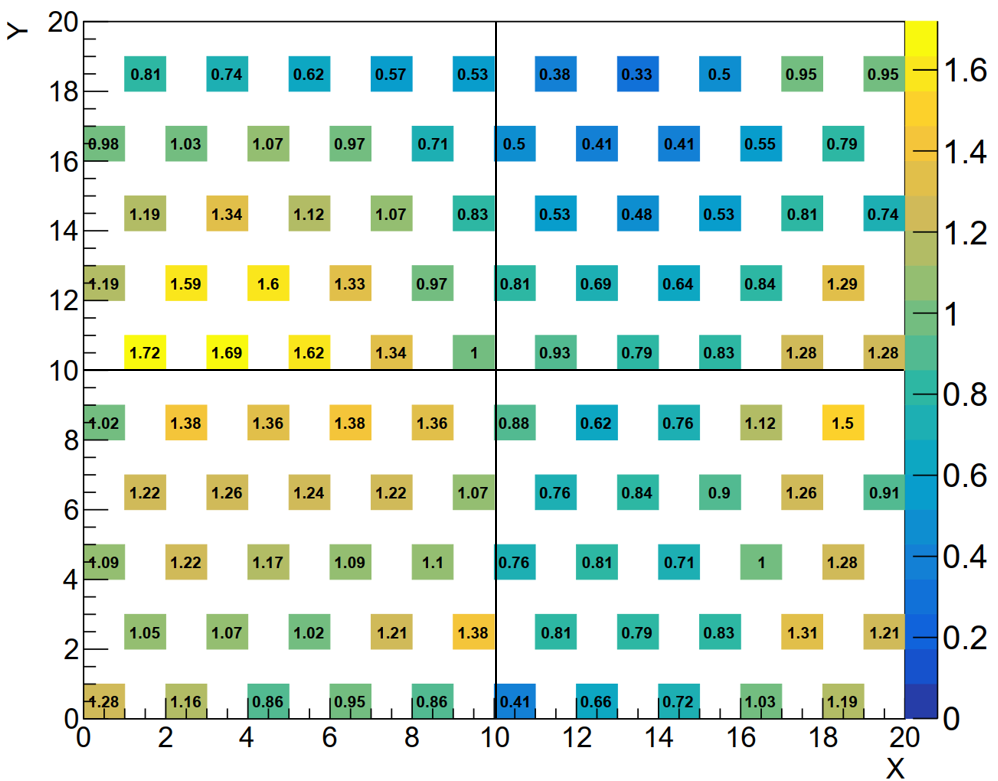} % 插入图片
        \vspace{2mm} % 图片与文字间的垂直间距
        \textbf{(b)} % 添加文字标注 (b)
    \end{minipage}
\caption{Gain uniformity map of (a) the 10 $\times$ 10 $\text{cm}^2$ PICOSEC MM and (b) the 20 $\times$ 20 $\text{cm}^2$ PICOSEC MM. The values on the channels represent the gain normalized to that of the channel in the middle. The uniformity is determined by the ratio of the standard deviation to the mean value of each channel.\label{fig:f2}}
\end{figure}

A series of tests were conducted on the large-area PICOSEC MM prototypes to evaluate their performance in laboratory conditions and test beam. The gain uniformity of the detector was tested with the single photoelectron spectrum, and the uniformity map is presented in figure \ref{fig:f2} (a) and figure \ref{fig:f2} (b), respectively. The gain uniformity of the 10 $\times$ 10 $\text{cm}^2$ PICOSEC MM prototype was measured to be 29.6\%, indicating a notable improvement following efforts to enhance the MM board's planarity. Still, the slight deformation of the MM board remains the primary contributor to the residual non-uniformity, exhibiting a trend of smaller in the center and larger at the edges. For the 20 $\times$ 20 $\text{cm}^2$ PICOSEC MM prototype, a subset of channels were measured to assess the overall uniformity of the detector, which was found to be approximately 32.3\%. In addition to the board deformation, the tilt of the four crystals also emerged as a factor affecting uniformity, as evident from the trends observed in the figures.

\begin{figure}[htbp]
\centering
    % 第一张图片
    \begin{minipage}[t]{0.45\textwidth} % 设置图片宽度为页面宽度的45%
        \centering % 图片和文字居中
        \includegraphics[width=\textwidth]{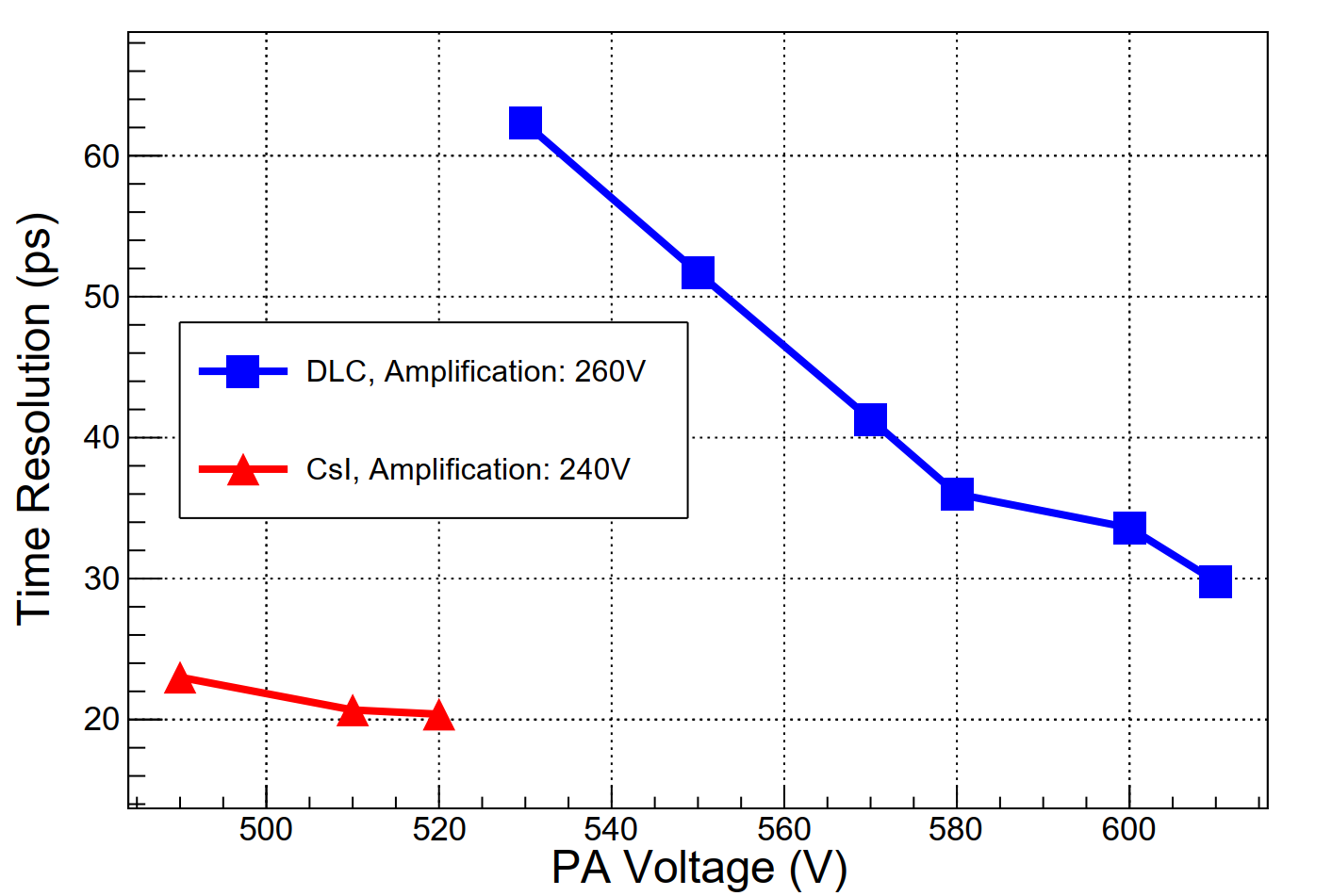} % 插入图片
        \vspace{2mm} % 图片与文字间的垂直间距
        \textbf{(a)} % 添加文字标注 (a)
    \end{minipage}
    \qquad
    % 第二张图片
    \begin{minipage}[t]{0.45\textwidth}
        \centering % 图片和文字居中
        \includegraphics[width=\textwidth]{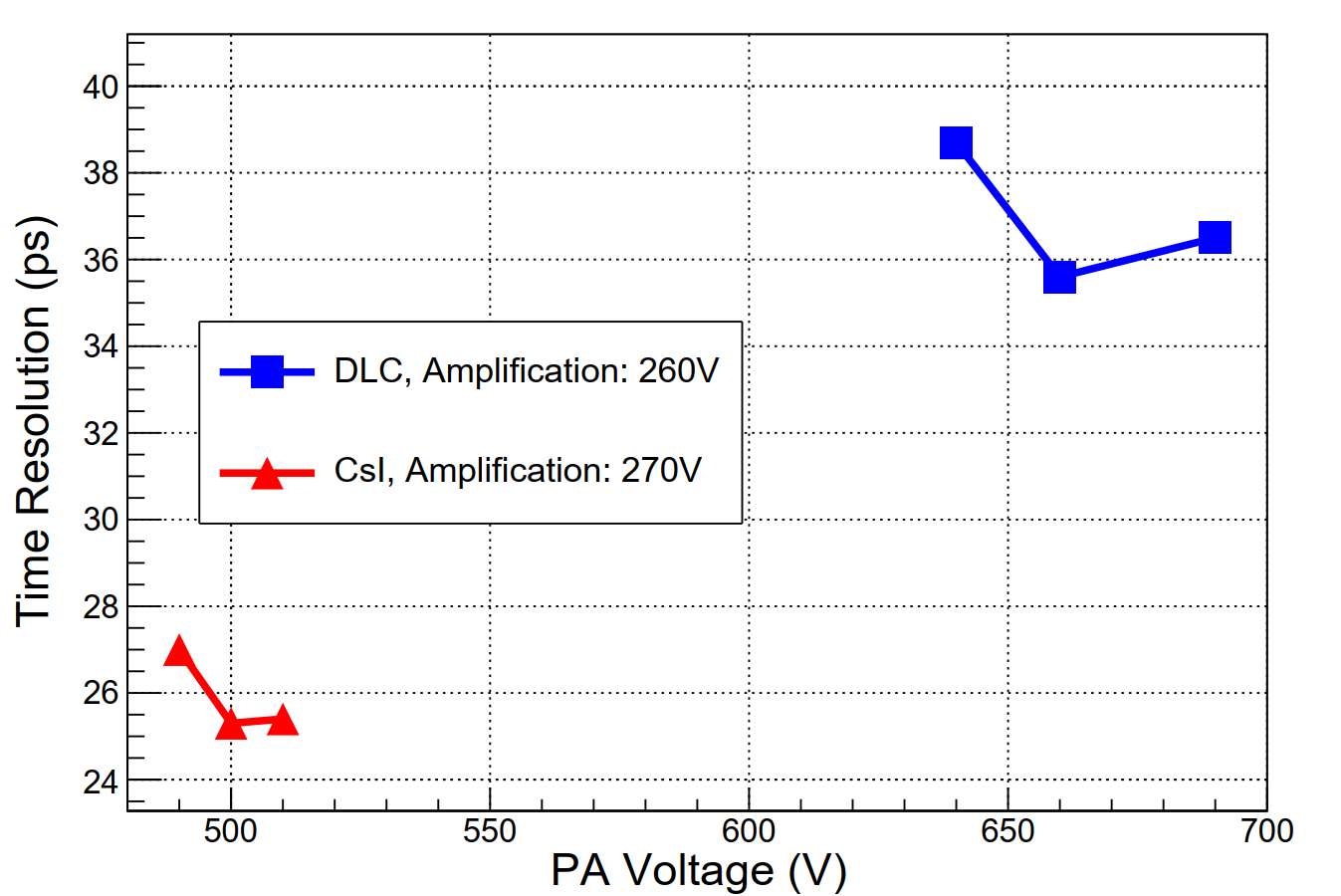} % 插入图片
        \vspace{2mm} % 图片与文字间的垂直间距
        \textbf{(b)} % 添加文字标注 (b)
    \end{minipage}
\caption{Time resolution results of (a) the 10 $\times$ 10 $\text{cm}^2$ PICOSEC MM prototype and (b) the 20 $\times$ 20 $\text{cm}^2$ PICOSEC MM prototype versus PA voltage with two different photocathodes: CsI and DLC.\label{fig:f3}}
\end{figure}

The time resolution of the prototypes was tested at the CERN SPS H4 beam line with 150 GeV/c muon beams, with the experimental setup detailed in \cite{sohl2020development}. Figure \ref{fig:f3} (a) shows the time resolution of the 10 $\times$ 10 $\text{cm}^2$ PICOSEC MM prototype as a function of the PA gap voltage with different photocathodes. The Cesium Iodide (CsI) photocathode, despite its relatively high yield of approximately ten photoelectrons (PEs) per Minimum Ionizing Particle (MIP), is susceptible to aging. In contrast, with the robustness characteristic, the DLC photocathode yields about three PEs per MIP \cite{Wang_2024}. With a CsI photocathode, the prototype achieved its optimal time resolution of 20.38 ps at the pad's central area, with PA and amplification voltages set to 520 V and 240 V, respectively. When equipped with a DLC photocathode, the prototype also achieved a time resolution of less than 30 ps, showing its excellent timing performance. Similar tests were conducted with the 20 $\times$ 20 $\text{cm}^2$ PICOSEC MM prototype to evaluate its time resolution. Figure \ref{fig:f3} (b) shows the time resolution of the prototype with different photocathodes tested across two different channels, both of which exhibit relatively higher gain. The detector equipped with a CsI photocathode achieved its optimal time resolution of 25 ps at the pad's center, with a PA voltage of 510 V and an amplification voltage of 210 V. This performance is comparable to that of the previous prototype, with the slightly degraded time resolution likely attributed to the thicker PA gap in this detector.

\section{Readout Electronics Development}
\begin{figure}[htbp]
\centering
    % 第一张图片
    \begin{minipage}[t]{0.75\textwidth} % 设置图片宽度为页面宽度的45%
        \centering % 图片和文字居中
        \includegraphics[width=\textwidth]{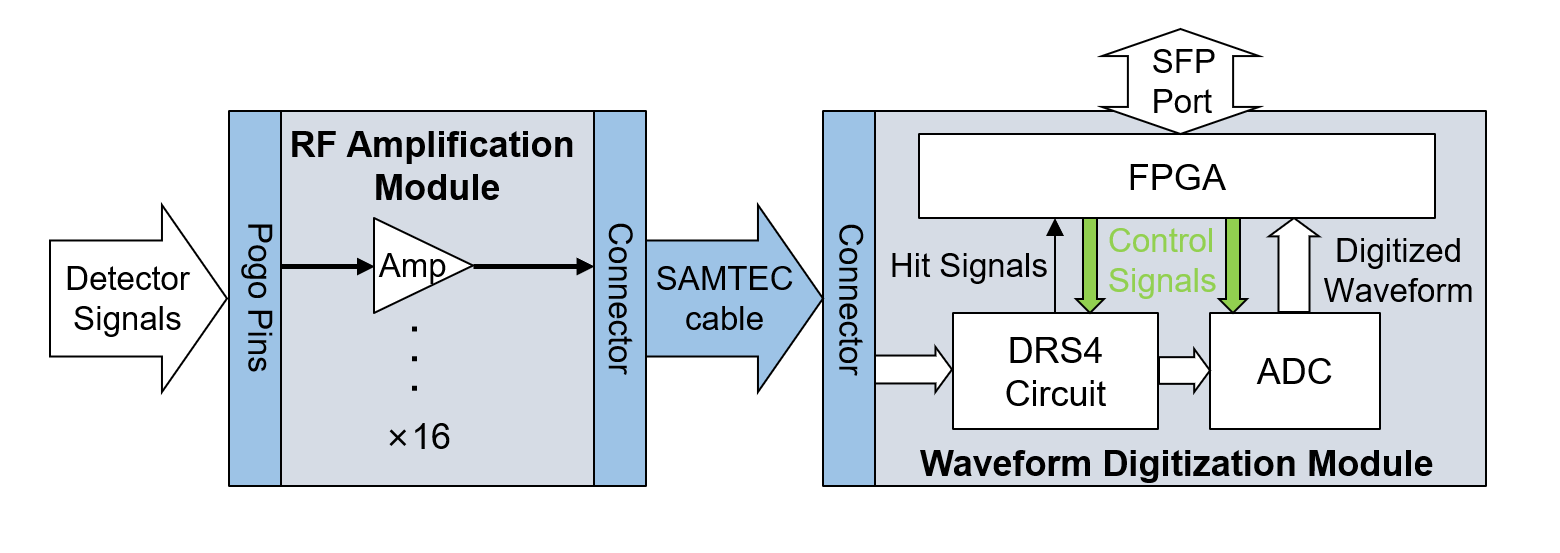} % 插入图片
        \vspace{2mm} % 图片与文字间的垂直间距
        \textbf{(a)} % 添加文字标注 (a)
    \end{minipage}
    \qquad
    % 第二张图片
    \begin{minipage}[t]{0.6\textwidth}
        \centering % 图片和文字居中
        \includegraphics[width=\textwidth]{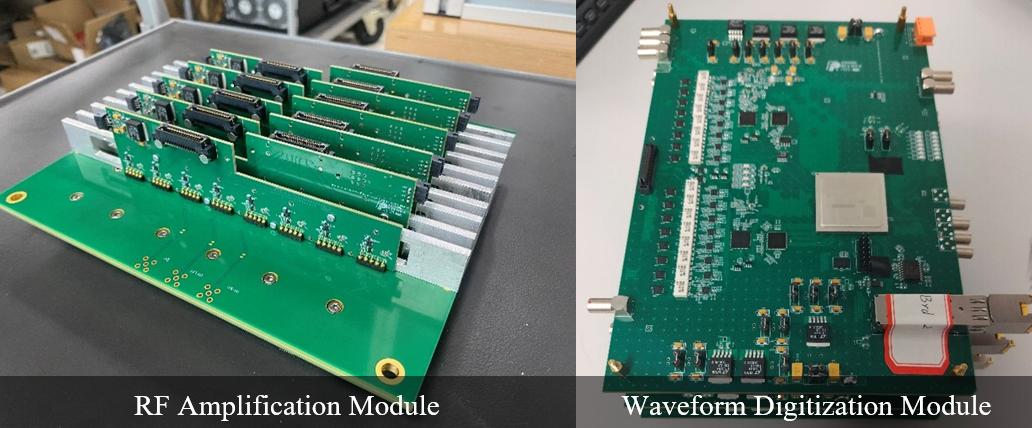} % 插入图片
        \vspace{2mm} % 图片与文字间的垂直间距
        \textbf{(b)} % 添加文字标注 (b)
    \end{minipage}
\caption{(a) Structure of the prototype readout electronics. (b) Photograph of the prototype readout electronics.\label{fig:f4}}
\end{figure}

\begin{figure}[htbp]
\centering
\includegraphics[width=.55\textwidth]{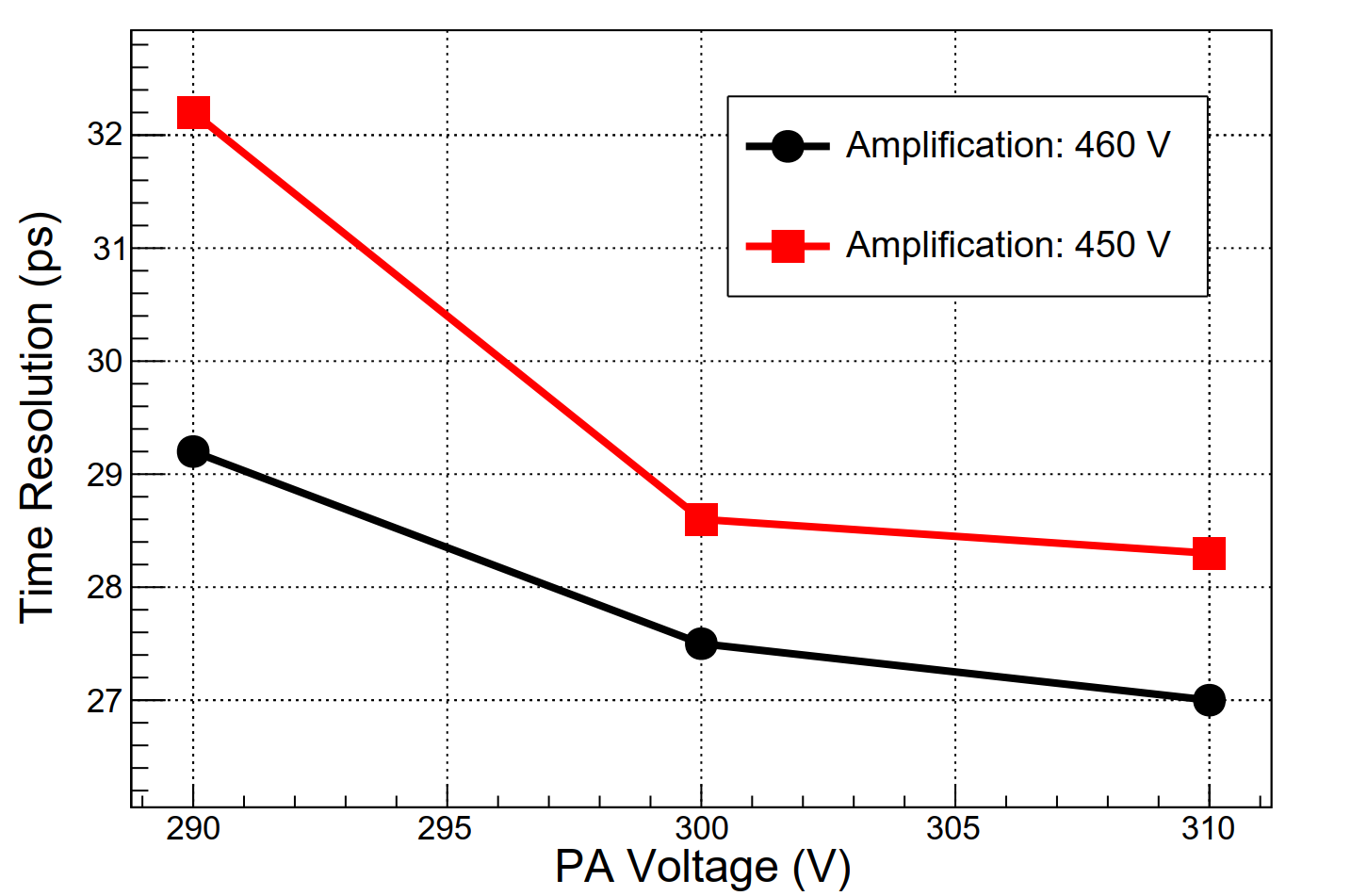}
\caption{Time resolution results of 20 $\times$ 20 $\text{cm}^2$ PICOSEC MM prototype in conjunction test with the readout electronics, equipped with a CsI photocathode. The results are shown for different amplification voltages and plotted as a function of PA voltage, with the optimal time resolution reaching below 30 ps.\label{fig:f5}}
\end{figure}

Customized readout electronics dedicated to high-precision time measurement for multi-channel PICOSEC were designed and tested. An electronic prototype consisting of an RF amplification module (RF-AM) and a Waveform Digitization Module (WDM) was designed, as illustrated in figure \ref{fig:f4}. The RF amplifiers, featuring 16 channels, are mechanically mounted on the Outer PCB of the PICOSEC MM to read out signals from either a 10 $\times$ 10 $\text{cm}^2$ or a 20 $\times$ 20 $\text{cm}^2$ PICOSEC MM prototype. Fast signals from the detectors are input via pogo pins to the RF-AM and then transmitted to the WDM through SAMTEC cables. After being sampled by the DRS4 chip and digitized by the Analog-to-Digital Converter (ADC), the signals are processed by the FPGA and subsequently uploaded to a PC via an SFP interface. During the aforementioned muon test beam, the entire system was tested in conjunction with the large-area PICOSEC MM prototypes. The performance of the RF-AM was individually tested while sampled by an oscilloscope to compare with a commercial amplifier Cividec\textsuperscript{1}\note{C2-HV Broadband amplifier (2 GHz, 40 dB), https://cividec.at/electronics-C2-HV.html}. The RF-AM demonstrated superior time resolution compared to the Cividec amplifier, due to its higher signal-to-noise ratio. Subsequently, the entire electronics was tested with the 20 $\times$ 20 $\text{cm}^2$ PICOSEC MM prototype equipped with a CsI photocathode, with the time resolution results presented in figure \ref{fig:f5}. The system possesses the ability to achieve an overall time resolution below 30 ps, which includes the combined time jitter contributions from the RF-AM, WDM, and the PICOSEC MM. These results highlight the system's high precision, integration, and reliability, making it well-suited for large-scale, multi-channel PICOSEC MM applications.

%The RF amplifier provides a gain of approximately 33 dB with a -3 dB bandwidth of 700 MHz. The WDM is based on a DRS4 chip operating at 5.12 Gsps and features a bandwidth of 950MHz.

Other fast-timing response electronics approaches were developed and extensively tested in conjunction with the PICOSEC MM. For example, a readout chain incorporating an RF Pulse Amplifier with a SAMPIC-based digitizer was employed \cite{hoarau2021rf,breton2016measurements}. In the test beam, this electronic setup was implemented to read out multiple channels of a 10 $\times$ 10 $\text{cm}^2$ PICOSEC MM \cite{lisowska2023towards}, showing a good ability to analyze the uniformity of its signal response. More recently, an integrated readout circuit specially designed to process fast signals based on the Application-Specific Integrated Circuit (ASIC), namely FastIC (Fast Integrated Circuit), was developed. The FastIC is an 8-channel front-end ASIC, making it a suitable approach for multi-channel readout of the PICOSEC MM. Its capability to process fast signals from the PICOSEC MM was validated in beam tests, where a time resolution of approximately 50 ps was achieved. Multi-channel readout of the PICOSEC MM using FastIC was subsequently tested, demonstrating its potential as a valuable option for the integrated readout of PICOSEC MM.

\section{Conclusion}
PICOSEC MM is undergoing continuous development aimed at detector optimization and large-area coverage. Several approaches to the 10 $\times$ 10 $\text{cm}^2$ resistive PICOSEC MM structure were developed, and a 20 $\times$ 20 $\text{cm}^2$ resistive PICOSEC MM was developed to expand the coverage by integrating four photocathodes. Beam tests with various photocathodes showed that the 10 $\times$ 10 $\text{cm}^2$ resistive PICOSEC MM achieved a time resolution of around 20 ps, while the 20 $\times$ 20 $\text{cm}^2$ PICOSEC MM reached a time resolution of 25 ps with a CsI photocathode, demonstrating excellent timing performance. Customized readout electronics dedicated to the multi-channel PICOSEC MM, including a readout chain of an RF-AM and WDM, were developed and thoroughly tested. The electronic system, when coupled with the large-area PICOSEC MM prototypes, achieved an overall time resolution below 30 ps during beam tests. This demonstrates that the PICOSEC MM and the entire system possesses the capability for high time precision and large-area integration, showcasing the potential for its application in future experiments with complex environments.

\acknowledgments

We acknowledge the support of the Program of National Natural Science Foundation of China (grant number 12125505); the CERN EP R\&D Strategic Programme on Technologies for Future Experiments; the RD51 collaboration, in the framework of RD51 common projects; the Cross-Disciplinary Program on Instrumentation and Detection of CEA, the French Alternative Energies and Atomic Energy Commission; the PHENIICS Doctoral School Program of Université Paris-Saclay, France; the COFUND-FP-CERN-2014 program (grant number 665779); the Fundação para a Ciência e a Tecnologia (FCT), Portugal; the Enhanced Eurotalents program (PCOFUND-GA-2013-600382); the US CMS program under DOE contract No. DE-AC02-07CH11359. This material is based upon work supported by the U.S. Department of Energy, Office of Science, Office of Nuclear Physics under contracts DE-AC05-06OR23177. The authors wish to thank the Hefei Comprehensive National Science Center for their support.

% Bibliography

%% [A] Recommended: using JHEP.bst file
%% \bibliographystyle{JHEP}
%% \bibliography{biblio.bib}

%% or
%% [B] Manual formatting (see below)
%% (i) We suggest to always provide author, title and journal data or doi:
%% in short all the informations that clearly identify a document.
%% (ii) please avoid comments such as "For a review'', "For some examples",
%% "and references therein" or move them in the text. In general, please leave only references in the bibliography and move all
%% accessory text in footnotes.
%% (iii) Also, please have only one work for each \bibitem.

% Bibliography

\bibliographystyle{JHEP}
 \bibliography{biblio.bib}
 
\end{document}